\RequirePackage{lineno}
\documentclass[aps,prl,twocolumn,superscriptaddress,showpacs,preprintnumbers,amsmath,amssymb,tightenlines]{revtex4}
\usepackage[dvipdfmx]{graphicx} % Include figure files
\usepackage{dcolumn}  % Align table columns on decimal point
\usepackage[dvipdfmx]{color}

\def\Vec#1{\mbox{\boldmath $#1$}}

\begin{document}

%\vspace*{-3\baselineskip}
%\resizebox{!}{3cm}{\includegraphics{belle.eps}}
%\linenumbers

\preprint{\vbox{ \hbox{   }
			\hbox{\today}
%			\hbox{\red{Ver.16}}
%			\hbox{Belle DRAFT {\it 2020-NN}}
%			\hbox{Belle Note 1358}
%			\hbox{Belle-CONF-1403}
%                       \hbox{Author: Y. Arita, K. Inami}
%                        \hbox{Committee: D. Epifanov(chair),}
%                        \hbox{T. Browder, H. Hayashii}
%  		              % \hbox{hep-ex nnnn}
}}

\title{ \quad\\[1.0cm]
An improved search for the electric dipole moment of the {\boldmath $\tau$} lepton
}

%%%% >>>>> insert the authorlist here. BEFORE the abstract !!!!! <<<<<
%%%% >>>>> from the authorship confirmation web page
%%% Name the file author.tex and use \input{author} to insert into your latex file.
%\author{K.~Inami}\affiliation{Nagoya university}
%\author{K.~Hayasaka}\affiliation{Niigata university}
%\collaboration{The Belle Collaboration}
%\noaffiliation
%% end author list
%%% Paper:    Tau EDM
%%% Journal:  Physical Review D
%%% Contacts: K. Inami (kenji@hepl.phys.nagoya-u.ac.jp)
%%% Non-responding authors or those who said NO are commented out.
%%% ====================================================================
%%% Click the RELOAD button on your web browser to see the updated file.
%%% ====================================================================
%%% Use \input{author} to insert this material into your latex file.
%%%%% Force institutions to appear in alphabetical order when typeset.
\noaffiliation
\affiliation{University of the Basque Country UPV/EHU, 48080 Bilbao}
%%%\affiliation{Beihang University, Beijing 100191}
\affiliation{University of Bonn, 53115 Bonn}
\affiliation{Brookhaven National Laboratory, Upton, New York 11973}
\affiliation{Budker Institute of Nuclear Physics SB RAS, Novosibirsk 630090}
\affiliation{Faculty of Mathematics and Physics, Charles University, 121 16 Prague}
%%%\affiliation{Chiba University, Chiba 263-8522}
\affiliation{Chonnam National University, Gwangju 61186}
\affiliation{University of Cincinnati, Cincinnati, Ohio 45221}
\affiliation{Deutsches Elektronen--Synchrotron, 22607 Hamburg}
%%%\affiliation{Duke University, Durham, North Carolina 27708}
\affiliation{University of Florida, Gainesville, Florida 32611}
%%%\affiliation{Department of Physics, Fu Jen Catholic University, Taipei 24205}
\affiliation{Key Laboratory of Nuclear Physics and Ion-beam Application (MOE) and Institute of Modern Physics, Fudan University, Shanghai 200443}
\affiliation{Justus-Liebig-Universit\"at Gie\ss{}en, 35392 Gie\ss{}en}
%%%\affiliation{Gifu University, Gifu 501-1193}
%%%\affiliation{II. Physikalisches Institut, Georg-August-Universit\"at G\"ottingen, 37073 G\"ottingen}
\affiliation{SOKENDAI (The Graduate University for Advanced Studies), Hayama 240-0193}
\affiliation{Gyeongsang National University, Jinju 52828}
\affiliation{Department of Physics and Institute of Natural Sciences, Hanyang University, Seoul 04763}
\affiliation{University of Hawaii, Honolulu, Hawaii 96822}
\affiliation{High Energy Accelerator Research Organization (KEK), Tsukuba 305-0801}
\affiliation{J-PARC Branch, KEK Theory Center, High Energy Accelerator Research Organization (KEK), Tsukuba 305-0801}
\affiliation{Higher School of Economics (HSE), Moscow 101000}
\affiliation{Forschungszentrum J\"{u}lich, 52425 J\"{u}lich}
%%%\affiliation{Hiroshima Institute of Technology, Hiroshima 731-5193}
\affiliation{IKERBASQUE, Basque Foundation for Science, 48013 Bilbao}
%%%\affiliation{University of Illinois at Urbana-Champaign, Urbana, Illinois 61801}
%%%\affiliation{Indian Institute of Science Education and Research Mohali, SAS Nagar, 140306}
\affiliation{Indian Institute of Technology Bhubaneswar, Satya Nagar 751007}
\affiliation{Indian Institute of Technology Guwahati, Assam 781039}
\affiliation{Indian Institute of Technology Hyderabad, Telangana 502285}
\affiliation{Indian Institute of Technology Madras, Chennai 600036}
\affiliation{Indiana University, Bloomington, Indiana 47408}
\affiliation{Institute of High Energy Physics, Chinese Academy of Sciences, Beijing 100049}
\affiliation{Institute of High Energy Physics, Vienna 1050}
\affiliation{Institute for High Energy Physics, Protvino 142281}
%%%\affiliation{Institute of Mathematical Sciences, Chennai 600113}
\affiliation{INFN - Sezione di Napoli, 80126 Napoli}
\affiliation{INFN - Sezione di Torino, 10125 Torino}
\affiliation{Advanced Science Research Center, Japan Atomic Energy Agency, Naka 319-1195}
\affiliation{J. Stefan Institute, 1000 Ljubljana}
%%%\affiliation{Kanagawa University, Yokohama 221-8686}
\affiliation{Institut f\"ur Experimentelle Teilchenphysik, Karlsruher Institut f\"ur Technologie, 76131 Karlsruhe}
\affiliation{Kavli Institute for the Physics and Mathematics of the Universe (WPI), University of Tokyo, Kashiwa 277-8583}
\affiliation{Kennesaw State University, Kennesaw, Georgia 30144}
%%%\affiliation{King Abdulaziz City for Science and Technology, Riyadh 11442}
\affiliation{Department of Physics, Faculty of Science, King Abdulaziz University, Jeddah 21589}
\affiliation{Kitasato University, Sagamihara 252-0373}
\affiliation{Korea Institute of Science and Technology Information, Daejeon 34141}
\affiliation{Korea University, Seoul 02841}
%%%\affiliation{Kyoto Sangyo University, Kyoto 603-8555}
%%%\affiliation{Kyoto University, Kyoto 606-8502}
\affiliation{Kyungpook National University, Daegu 41566}
\affiliation{Universit\'{e} Paris-Saclay, CNRS/IN2P3, IJCLab, 91405 Orsay}
%%%\affiliation{\'Ecole Polytechnique F\'ed\'erale de Lausanne (EPFL), Lausanne 1015}
\affiliation{P.N. Lebedev Physical Institute of the Russian Academy of Sciences, Moscow 119991}
\affiliation{Liaoning Normal University, Dalian 116029}
\affiliation{Faculty of Mathematics and Physics, University of Ljubljana, 1000 Ljubljana}
%%%\affiliation{Ludwig Maximilians University, 80539 Munich}
\affiliation{Luther College, Decorah, Iowa 52101}
\affiliation{Malaviya National Institute of Technology Jaipur, Jaipur 302017}
%%%\affiliation{University of Malaya, 50603 Kuala Lumpur}
\affiliation{University of Maribor, 2000 Maribor}
\affiliation{Max-Planck-Institut f\"ur Physik, 80805 M\"unchen}
\affiliation{School of Physics, University of Melbourne, Victoria 3010}
\affiliation{University of Mississippi, University, Mississippi 38677}
\affiliation{University of Miyazaki, Miyazaki 889-2192}
\affiliation{Moscow Physical Engineering Institute, Moscow 115409}
\affiliation{Graduate School of Science, Nagoya University, Nagoya 464-8602}
\affiliation{Kobayashi-Maskawa Institute, Nagoya University, Nagoya 464-8602}
\affiliation{Universit\`{a} di Napoli Federico II, 80126 Napoli}
%%%\affiliation{Nara University of Education, Nara 630-8528}
\affiliation{Nara Women's University, Nara 630-8506}
%%%\affiliation{National Central University, Chung-li 32054}
\affiliation{National United University, Miao Li 36003}
\affiliation{Department of Physics, National Taiwan University, Taipei 10617}
\affiliation{H. Niewodniczanski Institute of Nuclear Physics, Krakow 31-342}
\affiliation{Nippon Dental University, Niigata 951-8580}
\affiliation{Niigata University, Niigata 950-2181}
%%%\affiliation{University of Nova Gorica, 5000 Nova Gorica}
\affiliation{Novosibirsk State University, Novosibirsk 630090}
%%%\affiliation{Okinawa Institute of Science and Technology, Okinawa 904-0495}
\affiliation{Osaka City University, Osaka 558-8585}
%%%\affiliation{Osaka University, Osaka 565-0871}
\affiliation{Pacific Northwest National Laboratory, Richland, Washington 99352}
\affiliation{Panjab University, Chandigarh 160014}
\affiliation{Peking University, Beijing 100871}
\affiliation{University of Pittsburgh, Pittsburgh, Pennsylvania 15260}
\affiliation{Punjab Agricultural University, Ludhiana 141004}
%%%\affiliation{Research Center for Electron Photon Science, Tohoku University, Sendai 980-8578}
\affiliation{Research Center for Nuclear Physics, Osaka University, Osaka 567-0047}
\affiliation{Meson Science Laboratory, Cluster for Pioneering Research, RIKEN, Saitama 351-0198}
%%%\affiliation{Theoretical Research Division, Nishina Center, RIKEN, Saitama 351-0198}
%%%\affiliation{RIKEN BNL Research Center, Upton, New York 11973}
%%%\affiliation{Saga University, Saga 840-8502}
\affiliation{Department of Modern Physics and State Key Laboratory of Particle Detection and Electronics, University of Science and Technology of China, Hefei 230026}
\affiliation{Seoul National University, Seoul 08826}
\affiliation{Showa Pharmaceutical University, Tokyo 194-8543}
\affiliation{Soochow University, Suzhou 215006}
\affiliation{Soongsil University, Seoul 06978}
%%%\affiliation{University of South Carolina, Columbia, South Carolina 29208}
%%%\affiliation{Stefan Meyer Institute for Subatomic Physics, Vienna 1090}
\affiliation{Sungkyunkwan University, Suwon 16419}
\affiliation{School of Physics, University of Sydney, New South Wales 2006}
\affiliation{Department of Physics, Faculty of Science, University of Tabuk, Tabuk 71451}
\affiliation{Tata Institute of Fundamental Research, Mumbai 400005}
%%%\affiliation{Excellence Cluster Universe, Technische Universit\"at M\"unchen, 85748 Garching}
\affiliation{Department of Physics, Technische Universit\"at M\"unchen, 85748 Garching}
\affiliation{School of Physics and Astronomy, Tel Aviv University, Tel Aviv 69978}
\affiliation{Toho University, Funabashi 274-8510}
\affiliation{Department of Physics, Tohoku University, Sendai 980-8578}
\affiliation{Earthquake Research Institute, University of Tokyo, Tokyo 113-0032}
\affiliation{Department of Physics, University of Tokyo, Tokyo 113-0033}
\affiliation{Tokyo Institute of Technology, Tokyo 152-8550}
%%%\affiliation{Tokyo Metropolitan University, Tokyo 192-0397}
%%%\affiliation{Tokyo University of Agriculture and Technology, Tokyo 184-8588}
\affiliation{Utkal University, Bhubaneswar 751004}
\affiliation{Virginia Polytechnic Institute and State University, Blacksburg, Virginia 24061}
\affiliation{Wayne State University, Detroit, Michigan 48202}
\affiliation{Yamagata University, Yamagata 990-8560}
\affiliation{Yonsei University, Seoul 03722}

  \author{K.~Inami}\affiliation{Graduate School of Science, Nagoya University, Nagoya 464-8602} % Nagoya
  \author{K.~Hayasaka}\affiliation{Niigata University, Niigata 950-2181} % Niigata

% \author{A.~Abdesselam}\affiliation{Department of Physics, Faculty of Science, University of Tabuk, Tabuk 71451} % Tabuk
  \author{I.~Adachi}\affiliation{High Energy Accelerator Research Organization (KEK), Tsukuba 305-0801}\affiliation{SOKENDAI (The Graduate University for Advanced Studies), Hayama 240-0193} % KEK
% \author{K.~Adamczyk}\affiliation{H. Niewodniczanski Institute of Nuclear Physics, Krakow 31-342} % Krakow
% \author{J.~K.~Ahn}\affiliation{Korea University, Seoul 02841} % Korea
  \author{H.~Aihara}\affiliation{Department of Physics, University of Tokyo, Tokyo 113-0033} % Tokyo
  \author{S.~Al~Said}\affiliation{Department of Physics, Faculty of Science, University of Tabuk, Tabuk 71451}\affiliation{Department of Physics, Faculty of Science, King Abdulaziz University, Jeddah 21589} % Tabuk
% \author{K.~Arinstein}\affiliation{Budker Institute of Nuclear Physics SB RAS, Novosibirsk 630090}\affiliation{Novosibirsk State University, Novosibirsk 630090} % BINP
% \author{Y.~Arita}\affiliation{Graduate School of Science, Nagoya University, Nagoya 464-8602} % Nagoya
  \author{D.~M.~Asner}\affiliation{Brookhaven National Laboratory, Upton, New York 11973} % BNL
% \author{H.~Atmacan}\affiliation{University of Cincinnati, Cincinnati, Ohio 45221} % Cincinnati
  \author{V.~Aulchenko}\affiliation{Budker Institute of Nuclear Physics SB RAS, Novosibirsk 630090}\affiliation{Novosibirsk State University, Novosibirsk 630090} % BINP
  \author{T.~Aushev}\affiliation{Higher School of Economics (HSE), Moscow 101000} % HSE
  \author{R.~Ayad}\affiliation{Department of Physics, Faculty of Science, University of Tabuk, Tabuk 71451} % Tabuk
% \author{T.~Aziz}\affiliation{Tata Institute of Fundamental Research, Mumbai 400005} % Tata
  \author{V.~Babu}\affiliation{Deutsches Elektronen--Synchrotron, 22607 Hamburg} % DESY
% ...
% Lebedev
% \author{V.~Zhulanov}\affiliation{Budker Institute of Nuclear Physics SB RAS, Novosibirsk 630090}\affiliation{Novosibirsk State University, Novosibirsk 630090} % BINP
% \author{T.~Zivko}\affiliation{J. Stefan Institute, 1000 Ljubljana} % Ljubljana
% \author{A.~Zupanc}\affiliation{Faculty of Mathematics and Physics, University of Ljubljana, 1000 Ljubljana}\affiliation{J. Stefan Institute, 1000 Ljubljana} % Ljubljana
% \author{N.~Zwahlen}\affiliation{\'Ecole Polytechnique F\'ed\'erale de Lausanne (EPFL), Lausanne 1015} % Lausanne
\collaboration{The Belle Collaboration}

\begin{abstract}
We report a measurement of the electric dipole moment of the $\tau$ lepton
($d_\tau$) using an 833~fb$^{-1}$ data sample collected near
the $\Upsilon(4S)$ resonance,
with the Belle detector at the KEKB asymmetric-energy $e^+ e^-$
collider.
Using an optimal observable method, we obtain
the real and imaginary parts of $d_\tau$ as
${\rm Re}(d_\tau) = ( -0.62 \pm 0.63 ) \times 10^{-17} ~e{\rm cm}$ and
${\rm Im}(d_\tau) = ( -0.40 \pm 0.32 ) \times 10^{-17} ~e{\rm cm}$,
respectively.
These results are consistent with null electric dipole moment at the present
level of experimental sensitivity and improve 
the sensitivity by about a factor of three.
\end{abstract}


\maketitle

%%%% >>>> keep the final version single-spaced
%\tighten

{\renewcommand{\thefootnote}{\fnsymbol{footnote}}}
\setcounter{footnote}{0}

%\section{Introduction}

%Motivation, Previous analyses,
The electric dipole moment (EDM) of the $\tau$ lepton is a 
fundamental parameter that parameterizes 
time-reversal ({\it T}) or charge-conjugation--parity ({\it CP})
violation at the $\gamma\tau\tau$ vertex. 
In the Standard Model (SM),
{\it CP} violation arises due to an irreducible phase 
in the CKM matrix~\cite{ref:KM}, which predicts an unobservably small
$\tau$-lepton EDM ($d_\tau$) 
of order $10^{-37}$ $e$cm~\cite{ref:SMprediction}. 
Hence, observation of a nonzero $d_\tau$ value would be
a clear sign of new physics. Some
new physics models indicate a larger
EDM of order $10^{-19}$ $e$cm~\cite{ref:newphysics}.

The most sensitive previous measurement set an upper limit on the EDM of
order $10^{-17}$ $e$cm~\cite{ref:BellePrev};
the results were obtained by the Belle collaboration~\cite{ref:Belle}
using 29.5~fb$^{-1}$ of data collected at the KEKB collider~\cite{ref:KEKB} at 
a center-of-mass (CM) energy $\sqrt{s} = 10.58$~GeV.
The obtained real and imaginary parts of $d_\tau$ were
${\rm Re}(d_\tau) = (1.15 \pm 1.70) \times 10^{-17} ~e{\rm cm}$ and
${\rm Im}(d_\tau) = (-0.83 \pm 0.86) \times 10^{-17} ~e{\rm cm}$,
respectively. The corresponding limits were
$-2.2 \times 10^{-17}<{\rm Re}(d_\tau)<4.5 \times 10^{-17} ~e{\rm cm}$ and
$-2.5 \times 10^{-17}<{\rm Im}(d_\tau)<0.8 \times 10^{-17} ~e{\rm cm}$.

In this paper, we 
present updated results on $d_\tau$ using a much larger 
sample of 833~fb$^{-1}$ Belle data, of which
571~fb$^{-1}$ collected at the $\Upsilon(4S)$ resonance;
74~fb$^{-1}$ collected 60 MeV below it; and
188~fb$^{-1}$ collected near the $\Upsilon(1S)$, $\Upsilon(2S)$, 
$\Upsilon(3S)$, and $\Upsilon(5S)$ resonances. 
These samples are independent from 
the one used in the previous Belle result.
The sensitivity for Re($d_\tau$) and Im($d_\tau$) has improved by
about a factor of three, due to the increase of the data statistics
and improved analysis strategy.

%EDM effect, matrix elements
The effective Lagrangian for $\tau$-pair production including
the EDM term in the vertex is
\begin{equation}
{\cal L}= \bar{\tau} [ -eQ \gamma^\mu A_\mu -i d_\tau \sigma^{\mu\nu}
\gamma_5 \partial_\mu A_\nu] \tau.
\end{equation}
Including the EDM term, the
squared spin-density matrix ($\chi_{\rm prod}$) for the production
vertex 
in the process $e^+ e^- \to \tau^+ \tau^-$ is given by~\cite{ref:EDM}
\begin{equation}
\chi_{\rm prod} = \chi_{\rm SM} 
 + {\rm Re}(d_\tau) \chi_{\rm Re}
 + {\rm Im}(d_\tau) \chi_{\rm Im}
 + |d_\tau|^2 \chi_{d^2} ,
\label{eq:matrix}
\end{equation}
where $\chi_{\rm SM}$ is the SM term,
and $\chi_{\rm Re}$ and $\chi_{\rm Im}$ are the interference terms
between the SM and the EDM for the real and imaginary parts of $d_\tau$.
Here, $\chi_{d^2}$ is a higher-order EDM term,
which we can neglect since $d_\tau$ is small.
The matrix elements in Eq.~(\ref{eq:matrix})
%($\chi_{\rm SM}$, $\chi_{\rm Re}$, $\chi_{\rm Im}$, $\chi_{d^2}$) 
can be expressed using the momenta of the electron beam and 
the $\tau$ lepton,
and the spins of $\tau^+$ and $\tau^-$ in the $e^+e^-$ CM frame.
The interference terms are proportional to {\it CP}-odd 
spin-momentum correlation
terms
\begin{eqnarray}
\chi_{\rm Re} &\propto& 
 - \{ m_\tau + (k_0-m_\tau)(\hat{\Vec{k}}\cdot\hat{\Vec{p}})^2 \}
   (\Vec{S}_+ \times \Vec{S}_-)\cdot\hat{\Vec{k}} \nonumber \\
& &  + k_0 (\hat{\Vec{k}}\cdot\hat{\Vec{p}})
      (\Vec{S}_+ \times \Vec{S}_-)\cdot\hat{\Vec{p}}, 
\label{eq:corr:Mre} \\
\chi_{\rm Im} &\propto& 
 - \{ m_\tau + (k_0-m_\tau)(\hat{\Vec{k}}\cdot\hat{\Vec{p}})^2 \}
   (\Vec{S}_+ - \Vec{S}_-)\cdot\hat{\Vec{k}} \nonumber \\
& &  + k_0 (\hat{\Vec{k}}\cdot\hat{\Vec{p}})
      (\Vec{S}_+ - \Vec{S}_-)\cdot\hat{\Vec{p}}, 
\label{eq:corr:Mim}
\end{eqnarray}
where 
$k_0$ is the energy of the $\tau^\pm$,
$m_\tau$ is the $\tau$ mass,
$\Vec{p}$ is the three-momentum of the $e^+$,
$\Vec{k}$ is the three-momentum of the $\tau^+$,
$\Vec{S}_\pm$ are the spin vectors for the $\tau^\pm$,
and hats denote unit momenta.
In Eqs.~(\ref{eq:corr:Mre}) and (\ref{eq:corr:Mim}) above,
$\chi_{\rm Re}$ is {\it T}-odd and $\chi_{\rm Im}$ is {\it T}-even.
A more detailed discussion is given in Ref.~\cite{ref:BellePrev}.

Several {\it CP}-violating observables have been proposed in 
the literature~\cite{ref:newphysics}. For this study,
we use the so-called optimal observable method~\cite{ref:Optimal}
to obtain the $d_\tau$ values.
The optimal observables are
\begin{equation}
{\cal O}_{\rm Re} = \frac{\chi_{\rm Re}}{\chi_{\rm SM}},~~~
{\cal O}_{\rm Im} = \frac{\chi_{\rm Im}}{\chi_{\rm SM}}.
\end{equation}
%We calculate these observables on an event-by-event basis as 
%explained below.
They maximize sensitivity to the $\tau$ EDM.
%The elements, $\chi_{\rm SM/Re/Im}$, can be calculated using the momenta of
%recontructed $\tau$ decay products, as discussed later.
%From the definition of
%$\chi_{\rm Re}$ and $\chi_{\rm Im}$, 
%${\cal O}_{\rm Re}$ is the {\it CP}-odd/{\it T}-odd observable, and
%${\cal O}_{\rm Im}$ is the {\it CP}-odd/{\it T}-even observable.
The mean values of these observables 
($\langle {\cal O}_{\rm Re} \rangle, \langle {\cal O}_{\rm Im} \rangle$)
are linearly dependent on Re($d_\tau$) and Im($d_\tau$), 
\begin{equation}
 \langle {\cal O}_{\rm Re} \rangle = a_{\rm Re} {\rm Re}(d_\tau) + b_{\rm Re}, ~
 \langle {\cal O}_{\rm Im} \rangle = a_{\rm Im} {\rm Im}(d_\tau) + b_{\rm Im},
 \label{eq:relation0}
\end{equation}
since
\begin{eqnarray}
\langle{\cal O}_{\rm Re}\rangle &\propto&
\int {\cal O}_{\rm Re} \chi_{\rm prod} d\phi \nonumber\\
& = & \int \chi_{\rm Re} d\phi
+ {\rm Re}(d_\tau) \int \frac{(\chi_{\rm Re})^2}{\chi_{\rm SM}} d\phi,
\label{eq:obs}
\end{eqnarray}
where the integration is performed over the available phase space $\phi$ 
and
\begin{equation}
a_{\rm Re} = \int \frac{(\chi_{\rm Re})^2}{\chi_{\rm SM}} d\phi,
~~ b_{\rm Re} = \int \chi_{\rm Re} d\phi.
\label{eq:param}
\end{equation}
The expression for ${\cal O}_{\rm Im}$ is identical with the exchange of
``Re'' and ``Im'' in Eqs.~(\ref{eq:obs}) and (\ref{eq:param}).
The cross-term containing the integral of the product of $\chi_{\rm Re}$
and $\chi_{\rm Im}$ drops out because of their different symmetry properties.
To determine the coefficients, we have performed the integration
using Monte Carlo (MC) samples in order to account for detector effects.
In principle, the constant term ($b_{\rm Re/Im}$) should be zero 
as $\chi_{\rm Re/Im}$ is symmetric, but can be nonzero
owing to nonuniform 
acceptance of the detector. We therefore add this term.
Using linear relations, Re($d_\tau$) and Im($d_\tau$)
can be obtained from the measured 
values of $\langle{\cal O}_{\rm Re/Im}\rangle$.

%\section{Data}

We have used the data collected by the Belle detector 
for this analysis.
%collected at the 
%$\Upsilon(4S)$ resonanc%Although the $\tau$ flight direction is necessary to calculate 
%the spin density matrix and the observables~\cite{ref:OPAL},
%e (636.4~fb$^{-1}$), 
%$\Upsilon(1S)$ (7.6~fb$^{-1}$), 
%$\Upsilon(2S)$ (26.6~fb$^{-1}$), 
%$\Upsilon(3S)$ (3.2~fb$^{-1}$), 
%$\Upsilon(5S)$ (122.8~fb$^{-1}$) and in an
%energy scan around the resonances (27.8~fb$^{-1}$).
%In total, we use 824.5~fb$^{-1}$ of data.
Belle is a large-solid-angle magnetic
spectrometer that consists of a silicon vertex detector,
a 50-layer central drift chamber (CDC), an array of
aerogel threshold Cherenkov counters (ACC),  % <- \v{C}erenkov 2007.08
a barrel-like arrangement of time-of-flight
scintillation counters (TOF), and an electromagnetic calorimeter (ECL)
comprised of CsI(Tl) crystals; all located inside 
a superconducting solenoid coil that provides a 1.5~T
magnetic field.  An iron flux-return yoke located outside of
the coil is instrumented to detect $K_L^0$ mesons and muons (KLM).  
The detector is described in detail elsewhere~\cite{ref:Belle}.

The MC event generators KKMC and TAUOLA~\cite{ref:KKMC}
are used for $\tau$-pair production and decays, respectively.
Detector simulation is performed by a GEANT3~\cite{ref:GEANT} based program.
We use a sample of MC events corresponding to about five times the
data luminosity.
%a luminosity of 3.8 ab$^{-1}$.
In order to study the background contamination arising
from non $\tau$-pair events,
we generate MC samples for the 
$e^+e^- \to q \bar{q}$ ($q=u, d, c, s$) continuum and
$e^+e^- \to \Upsilon(4S) \to B \bar{B}$ events
using the EVTGEN~\cite{ref:EVTGEN} program,
and for two-photon mediated processes 
($e^+ e^- \to e^+ e^- \ell^+\ell^-, e^+ e^- q\bar{q}$)
using the AAFH~\cite{ref:AAFHB} program.

%Event selection
We use $\tau$-pair events with a 1-prong versus 1-prong topology
in which the particles are selected by the following criteria.
Charged tracks are required to have a transverse momentum of 
$p_{\mathrm T}>0.1$~GeV/$c$ and an impact parameter
%in the laboratory frame, and
%a distance of closest approach with respect to the interaction point
along the positron beam and in the transverse plane less then
3.0~cm and 1.0~cm, respectively.
An ECL cluster not matching any track is identified as a photon candidate.
Photon candidates should deposit an energy of %in the laboratory frame of 
$E>0.1$~GeV in the ECL.
Each charged particle is identified using a likelihood ratio formed
combining the ionization energy loss in the CDC, 
the ratio of energy deposited in the ECL
and momentum measured in the CDC, the shower shape in the ECL,
the position matching of the ECL cluster and CDC track,
the range and hit pattern in the KLM, the time-of-flight information 
from the TOF, and the light yield of the ACC.
%Electron, muon and pion candidates should satisfy
%stringent probability requirements.
%A kaon veto is also applied for pion selection.
Electron and muon candidates are selected by requiring the
likelihood ratios 
${\cal P}(e)$~\cite{ref:eid} and ${\cal P}(\mu)$~\cite{ref:muid}
to exceed 0.9 and 0.95, respectively.
The corresponding identification efficiency is above 90\%
with a pion misidentification rate less than 2\%.
Charged pions and kaons are distinguished using likelihood ratios,
${\cal P}(i/j) = {\cal L}_i / ({\cal L}_i +{\cal L}_j)$,
where ${\cal L}_i$ is the likelihood for a track to be identified as $i$.
Pion candidates for the $\tau \to \pi \nu$ mode are
selected by requiring ${\cal P}(K/\pi)<0.8$,
${\cal P}(\mu)<0.05$, ${\cal P}(e)<0.01$, and an electron likelihood ratio
obtained by combining information from the ACC and CDC less than 0.9
to reduce electron backgrounds, which do not interact in the ECL.
The requirement ${\cal P}(K/\pi)<0.8$ rejects 78\% of kaons, while
94\% of muons are rejected by the requirement ${\cal P}(\mu)<0.05$
and 98\% of electrons are rejected by the requirement ${\cal P}(e)<0.01$.
A $\rho^\pm$ is reconstructed from a charged track and a $\pi^0$,
requiring the opening angle between them
to be less than $90^\circ$ in the CM frame and
the charged track not to be an electron or a muon.
The $\rho^\pm$ candidates include higher $\rho$ resonances
since no mass cut is applied.
The $\pi^0$ candidates,
reconstructed from $\gamma\gamma$ combinations, should have an invariant
mass between 110 and 150 MeV/$c^2$ and a momentum of $p>0.2$~GeV/$c$.

We select eight exclusive final states of the $\tau$-pair process
$\tau \tau \to (e\nu\bar{\nu})(\mu\nu\bar{\nu})$, $(e\nu\bar{\nu})(\pi\nu)$, 
$(\mu\nu\bar{\nu})(\pi\nu)$, $(e\nu\bar{\nu})(\rho\nu)$, 
$(\mu\nu\bar{\nu})(\rho\nu)$, $(\pi\nu)(\rho\nu)$, 
$(\rho\nu)(\rho\bar{\nu})$, and $(\pi\nu)(\pi\bar{\nu})$.
Hereafter, we refer to these final states as $e\mu$, $e\pi$, $\mu\pi$, $e\rho$,
$\mu\rho$, $\pi\rho$, $\rho\rho$, and $\pi\pi$, respectively. 
We require two charged tracks with zero net charge
and no photons except for the daughters of the $\rho^\pm$ in each event.
The sum of the momenta of charged tracks and photons 
should be less than 9~GeV/$c$. 
(All kinematical values are defined in the laboratory frame, 
unless otherwise noted.)
In order to reduce the background and enhance the particle-identification 
separation power,
the lepton is required to lie within the barrel region, 
$-0.60<\cos \theta< 0.83$, 
while the $\pi^\pm$ is required to be within $-0.50<\cos \theta<0.62$,
where $\cos \theta$ is 
the cosine of the polar angle. % in the laboratory frame.
%Because of the large background hits in the KLM endcap region,
%the $\cos \theta$ region is narrow.
Furthermore, we require the momentum to
be greater than 0.5~GeV/$c$ for an electron,
1.2~GeV/$c$ for a muon or pion, and 1.0~GeV/$c$ for a $\rho^\pm$.

In order to suppress two-photon mediated background contributions,
we require the missing momentum vector not to point
along the beam pipe, $-0.950 < \cos \theta_{\rm miss}<0.985$.
%in the laboratory frame.
To reject Bhabha scattering and $\mu\mu$ backgrounds, we require the sum of
the charged track momenta in the CM frame be less than
9.0~GeV/$c$.
For the $e\pi$ mode, 
we remove events
if at least one of the following conditions is satisfied:
the opening angle between the two charged particles
in the plane perpendicular to the beam axis
is greater than $175^\circ$, the sum of the charged track
momenta in the CM frame is greater than 6.0~GeV/$c$,
or the $E_{\rm ECL}/p$ of pion is larger than 1.05. 
These criteria are required
because of the large contamination from radiative Bhabha events.
In addition,
for the $e\rho$ mode, we require that the electron momentum in the CM frame
be less than 5~GeV/$c$ to suppress the same background.

These selection criteria are similar to those required in the previous
analysis, with some changes following updates to the reconstruction software,
and updated detector calibration.

To maintain consistency between the data and simulation,
the effect of the trigger~\cite{ref:trg} should be taken into account.
A hardware trigger simulator is used for MC samples.
%The trigger efficiency is about 96\%.
We also reject events in which
the $\tau$ flight direction cannot be kinematically reconstructed
in the observable calculation discussed below.
%To meet this demand the many background processes and
%$\tau$-pairs with hard initial state radiation are rejected.

%We have applied the similar selection criteria with the previous analysis,
%though it was tuned following the updates of reconstruction software and
%the detector calibration.
%obtained data
The obtained signal yield, purity, and dominant background are listed in 
Table~\ref{table:selection.result}.
The purity and the background are estimated using MC samples.
In some modes, $\tau$ decays with additional
$\pi^0$ mesons contribute a significant background due to 
low energy photons that escape detection.
\begin{table*}[htb]
\caption{Yield, purity, and dominant backgrounds for each selected mode.
%The purity and background are evaluated using MC simulation.
The values in square brackets indicate the expected background rates in \%.
}
\label{table:selection.result}
\begin{tabular}{@{\hspace{0.3cm}}c@{\hspace{0.3cm}}r@{\hspace{0.3cm}}c@{\hspace{0.3cm}}l@{\hspace{0.3cm}}}
\hline\hline
Mode   & Yield~~~~ & Purity(\%) & Background (\%) \\
\hline
$e\mu$     & 6434268  & 95.8 & two-photon process ($ee\mu\mu)$ [2.5], $\tau\tau \to (e \nu\nu)(\pi\nu)$ [1.3]  \\
$e\pi$     & 2644971  & 85.7 & $\tau\tau \to (e\nu\nu)(\rho\nu)$ [6.5], $(e\nu\nu)(\mu\nu\nu)$ [5.1], $(e\nu\nu)(K^*\nu)$ [1.3] \\
$\mu\pi$   & 2503936  & 80.5 & $\tau\tau \to (\mu\nu\nu)(\rho\nu)$ [6.4], $(\mu\nu\nu)(\mu\nu\nu)$ [4.9], $(\mu\nu\nu)(K^*\nu)$ [1.3], two-photon process ($ee\mu\mu)$ [3.1] \\
$e\rho$    & 7218823  & 91.7 & $\tau\tau \to (e\nu\nu)(\pi\pi^0\pi^0\nu)$ [4.6], $(e\nu\nu)(K^*\nu)$ [1.7] \\
$\mu\rho$  & 6203489  & 91.0 & $\tau\tau \to (\mu\nu\nu)(\pi\pi^0\pi^0\nu)$ [4.3], $(\mu\nu\nu)(K^*\nu)$ [1.6], $(\pi\nu)(\rho\nu)$ [1.1] \\
$\pi\rho$  & 2655696  & 77.0 & $\tau\tau \to (\rho\nu)(\rho\nu)$ [6.7], $(\pi\nu)(\pi\pi^0\pi^0\nu)$ [3.9], $(\mu\nu\nu)(\rho\nu)$ [5.1], $(\rho\nu)(K^*\nu)$ [1.4], $(\pi\nu)(K^*\nu)$ [1.4] \\
$\rho\rho$ & 3277001 & 82.4 & $\tau\tau \to (\rho\nu)(\pi\pi^0\pi^0\nu)$ [9.4], $(\rho\nu)(K^*\nu)$ [3.1] \\
$\pi\pi$   & 460288  & 71.9 & $\tau\tau \to (\pi\nu)(\rho\nu)$ [11.3], $(\pi\nu)(\mu\nu\nu)$ [8.8], $(\pi\nu)(K^*\nu)$ [2.5] \\
\hline\hline
\end{tabular}
\end{table*}

\begin{figure}[htb]
\includegraphics[width=0.5\textwidth]{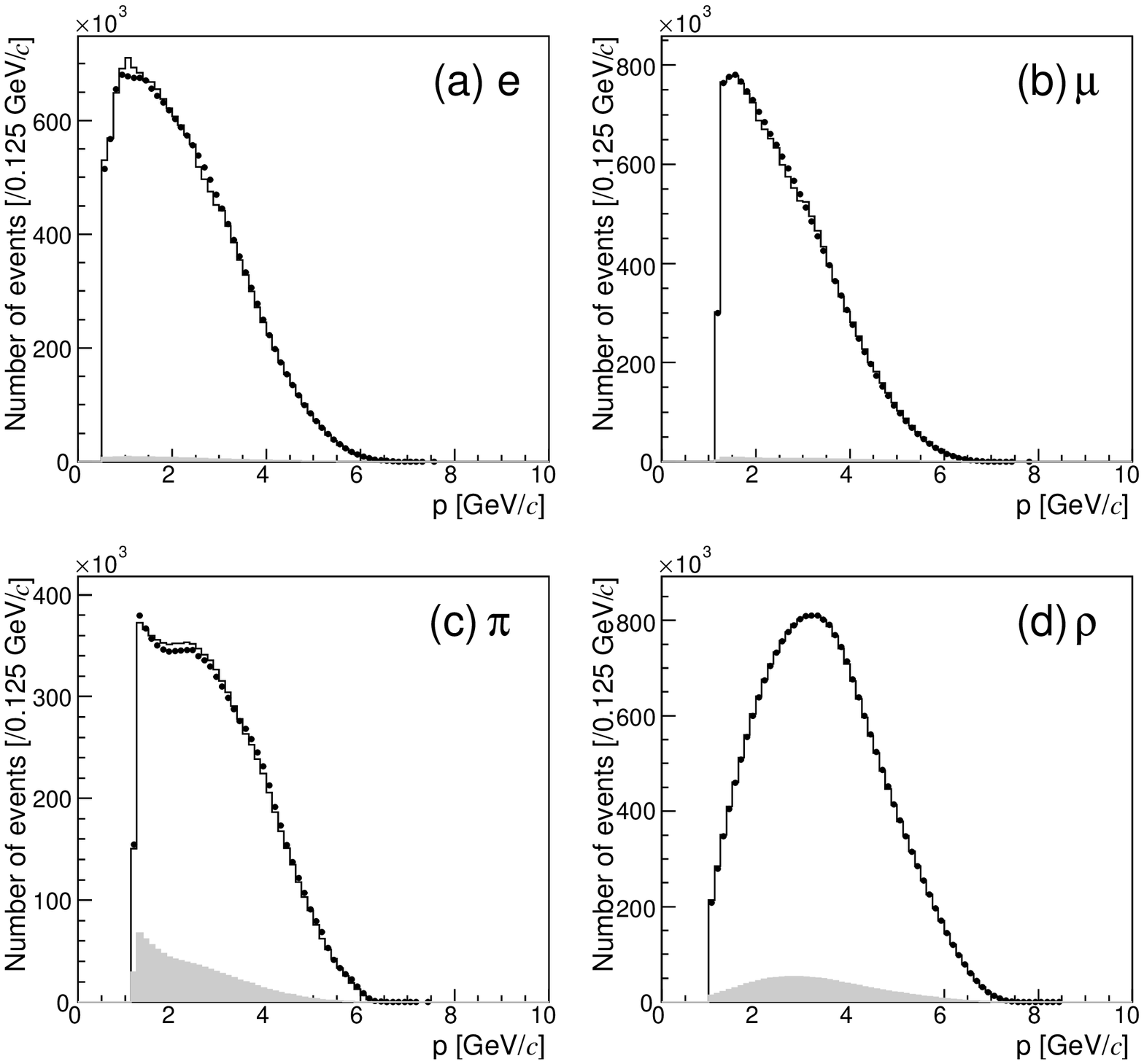}
\caption{Momentum distributions of (a) electrons, (b) muons, (c) pions,
and (d) $\rho$ mesons %in the laboratory frame,
for the samples obtained after all event selections in each mode.
The points with error bars are the data,
the solid histograms are the MC expectation, and the gray
shaded histograms are the contribution from misidentification
for each particle species.}
\label{fig:mom}
\end{figure}

\begin{figure}[htb]
\includegraphics[width=0.5\textwidth]{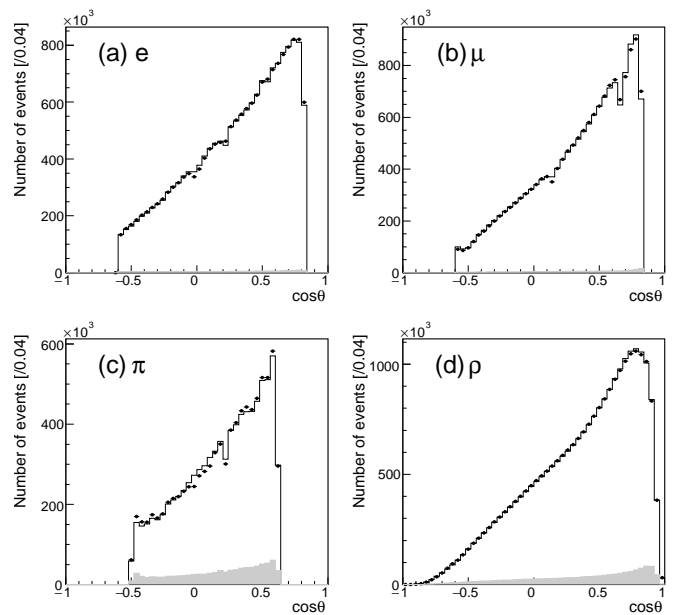}
\caption{The $\cos \theta$ distributions of (a) electrons, (b) muons, (c) pions,
and (d) $\rho$ mesons %in the laboratory frame,
for the samples obtained after all event selections in each mode.
The points with error bars are the data,
the solid histograms are the MC expectation, and the gray
shaded histograms are the contribution from misidentification
for each particle species.}
\label{fig:cos}
\end{figure}

The momentum and $\cos \theta$ distributions %in the laboratory frame
for the obtained samples are shown 
in Figs.~\ref{fig:mom} and \ref{fig:cos}.

%\section{Analysis}

%Although the $\tau$ flight direction is necessary to calculate 
%the spin density matrix and the observables~\cite{ref:OPAL},

\begin{figure*}[htb]
\includegraphics[width=0.9\textwidth]{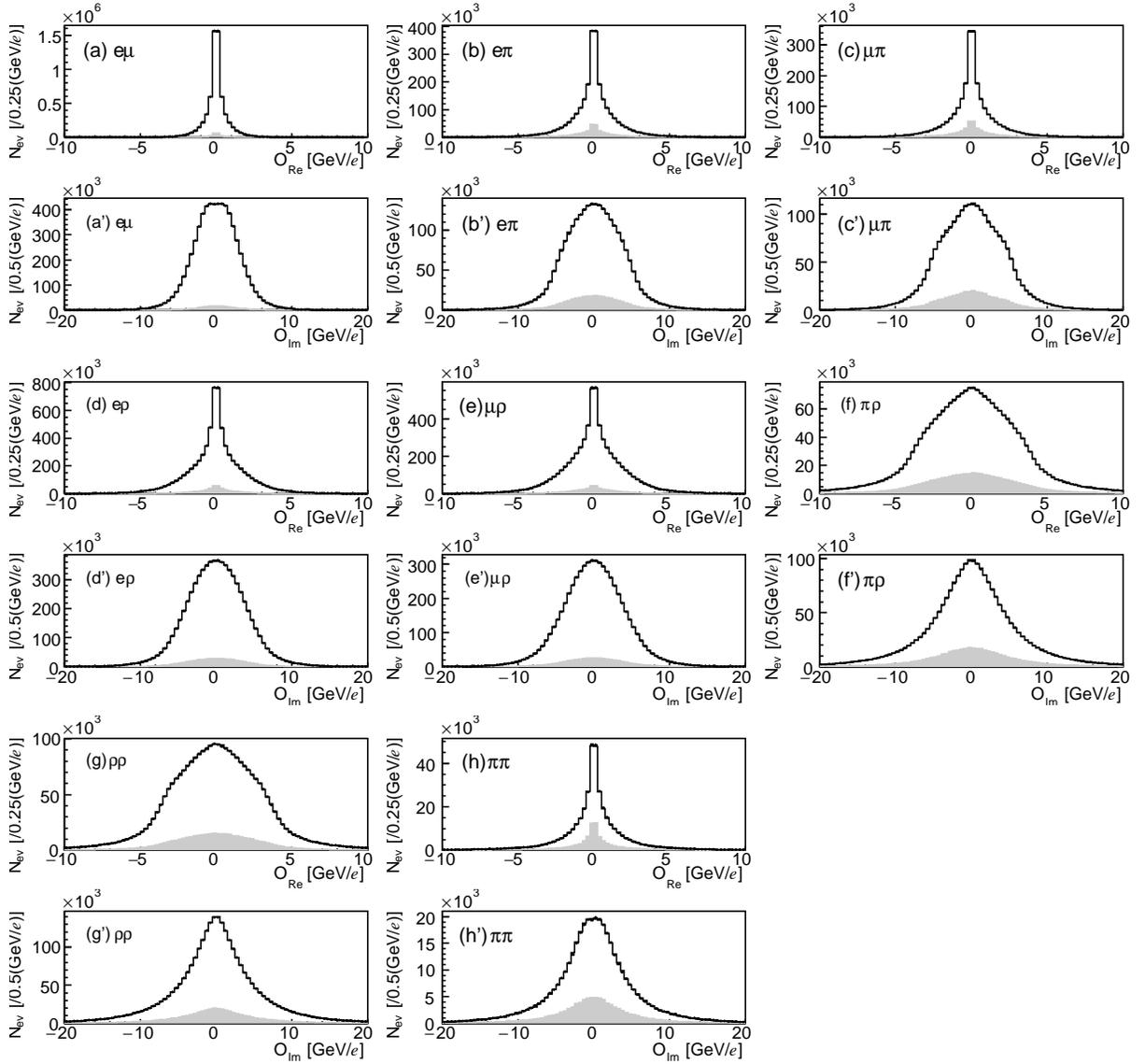}
\caption{Distributions of optimal observable for each mode.
The upper (a)--(h) plots are ${\cal O}_{\rm Re}$ 
and the lower (a')--(h') plots are ${\cal O}_{\rm Im}$ for each mode.
The points with error bars are the data and 
the solid histograms are the MC expectation with zero EDM.
The gray shaded histograms are the background contribution estimated from 
simulation.}
\label{fig:obs}
\end{figure*}

In order to calculate the observables, we need to determine the
$\tau$ spin vectors and flight direction.
The quantities used in the following calculation are obtained
in the CM frame.
The spin vectors,
which give the most probable direction of the spin,
are reconstructed using the momenta of $\tau$ and its 
decay products~\cite{ref:SpinVector,ref:OPAL}.
For example, the spin vectors for $\tau^\pm \to \pi^\pm \nu_\tau$ are given by
\begin{equation}
\Vec{S}_\pm = \frac{2}{m_\tau^2 - m_\pi^2}
\left( \mp m_\tau \Vec{p}_{\pi^\pm} + 
       \frac{m_\tau^2+m_\pi^2+2m_\tau E_{\pi^\pm}}{2(E_\tau + m_\tau)}\Vec{k}
\right),
\end{equation}
where $\Vec{p}_{\pi^\pm}$ and $E_{\pi^\pm}$ are the 
$\pi^\pm$ momentum and energy,
respectively. (See the Appendix for other decays.)
Although the $\tau$ flight direction $\hat{\Vec{k}}$ 
is necessary to calculate 
the spin vector and observables~\cite{ref:OPAL},
experimentally the $\tau$ direction cannot be uniquely determined
due to the presence of two or more missing neutrinos.
In the reactions where both $\tau$ leptons decay semileptonically,
$e^+e^- \to \tau^+\tau^- \to A^+B^-\nu_\tau\bar{\nu_\tau}$
without initial-state radiation (ISR),
the two possible solutions for
the unit vector of the $\tau^+$ flight direction, 
$\hat{\Vec{k}}_+$ and $\hat{\Vec{k}}_-$,
are given by
\begin{equation}
\hat{\Vec{k}}_{\pm} = u \hat{\Vec{p}}_A + v \hat{\Vec{p}}_B 
 \pm w \frac{\Vec{p}_A \times \Vec{p}_B}{|\Vec{p}_A \times \Vec{p}_B|},
\label{eq:taudirection}
\end{equation}
where $\Vec{p}_A$ ($\Vec{p}_B$) are the sum of three-momentum vectors in the
decay products, $A^+$ ($B^-$).
%and the hats denote unit momenta.
The parameters $u$, $v$, and $w$ are
\begin{eqnarray}
u &=& \frac{\cos \theta_A + \hat{\Vec{p}}_A \cdot\hat{\Vec{p}}_B \cos \theta_B}
 { 1- (\hat{\Vec{p}}_A \cdot\hat{\Vec{p}}_B)^2 }, \\
v &=& - \frac{\cos \theta_B + \hat{\Vec{p}}_A \cdot\hat{\Vec{p}}_B \cos \theta_A}
 { 1- (\hat{\Vec{p}}_A \cdot\hat{\Vec{p}}_B)^2 }, \\
w &=& \sqrt{1-u^2-v^2-2uv(\hat{\Vec{p}}_A \cdot\hat{\Vec{p}}_B)}, \label{eq:w}
\end{eqnarray}
where $\theta_A$ ($\theta_B$) are the angles between the momenta of
the decay product $A^+$ ($B^-$) and the $\tau$ momentum:
\begin{equation}
\cos \theta_i = \frac{2 E_\tau E_i-m^2_i-m^2_\tau}
 {2 |\Vec{k}||\Vec{p}_i|},
\label{eq:costheta}
\end{equation}
where $i = A$ or $B$.
%When both $\tau$ leptons decay to hadrons, 
In this case, the $\tau$ direction 
can be obtained with a twofold ambiguity.
Experimentally this ambiguity cannot be resolved. 
Therefore, we take an average of the 
two possible solutions in the calculation of the observables.
In the case of leptonic $\tau$ decays, one more ambiguity in
the invariant mass of two neutrinos from the same $\tau$, $m_{\nu\nu}$, 
arises as
\begin{equation}
\cos \theta_\ell = \frac{2 E_\tau E_\ell-m^2_\ell-m^2_\tau+m^2_{\nu\nu}}
 {2 |\Vec{k}||\Vec{p}_\ell|}.
\label{eq:costheta2}
\end{equation}
We then take an average over multiple solutions 
using the MC method by varying $m_{\nu\nu}$ uniformly within the 
possible kinematical range.
%observable calculation
%As discussed, 
%experimentally the $\tau$ direction cannot be obtained without any ambiguity,
%because two or more neutrinos are missed in a $\tau$-pair event.
%Therefore, for each event we must find the range of kinematic configurations, 
%which are consistent with the observed final state,
%and use the average value for the possible configurations.
%In the case where both $\tau$ leptons decay to hadrons ($\tau \to \pi \nu$ 
%or $\rho \nu$), the $\tau$ flight direction
%can be determined with the twofold ambiguity~\cite{ref:OPAL},
%and we calculate the average of the two solutions.
%In the case that one or both $\tau$ decay leptonically 
%($\tau \to e \nu\bar{\nu}$ and $\mu \nu\bar{\nu}$),
%we would use a Monte Carlo method.
%In this case, we have one more ambiguity due to the missing two neutrinos,
%where the invariant mass of the two neutrinos $m_{\nu\nu}$ is undermined.
For each event, we make 100 trials
using a ``hit-and-miss'' approach
while varying the effective mass $m_{\nu\nu}$ randomly.
With $N_{\rm hit}$ successful trials in which the $\tau$ direction can be
constructed kinematically, the average value of the
observable is obtained for each event.
In the case where both $\tau$'s decay leptonically, the $m_{\nu\nu}$ is varied
for each $\tau$.
In the calculation, we require $w$ in Eq.~(\ref{eq:w}) be real and 
$\cos \theta_j ~(j = A, B, \ell)$
in Eqs.~(\ref{eq:costheta}) and (\ref{eq:costheta2})
be within the range $[-1, 1]$, therefore we removed the cases
when the above requriements were not satisfied. 
In the analysis, we neglect the effect of ISR
for the calculation of the observables, and treat it
as a systematic source.
The distributions of observables for the obtained samples are shown in 
Fig.~\ref{fig:obs}, 
along with those obtained from MC simulations with no EDM.
We calculate the mean value of each observable using the data in the full
range including events beyond the range shown in Fig.~\ref{fig:obs}.

\begin{figure}[htb]
\includegraphics[width=0.35\textwidth]{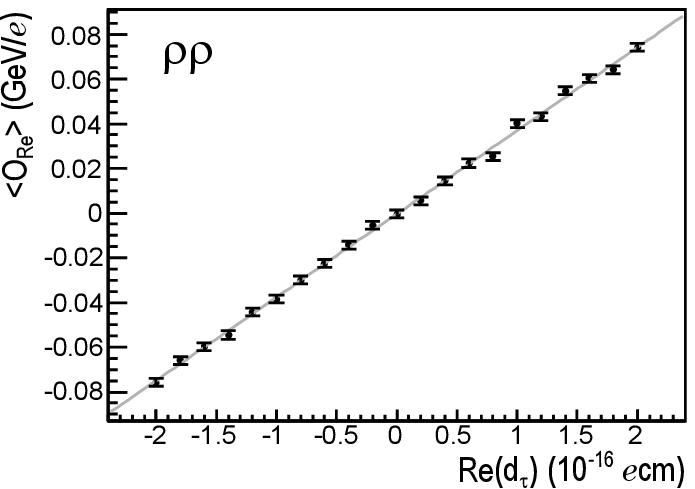}
\caption{Relation of Re($d_\tau$) and $\langle{\cal O}_{\rm Re}\rangle$
for the $\rho\rho$ mode obtained by the MC simulation.
The line shows the fitted function.
Other modes also show a similar linear dependence; the non-linearity
is negligible for all modes.}
\label{fig:relation}
\end{figure}

\begin{figure}[htb]
\includegraphics[width=0.45\textwidth]{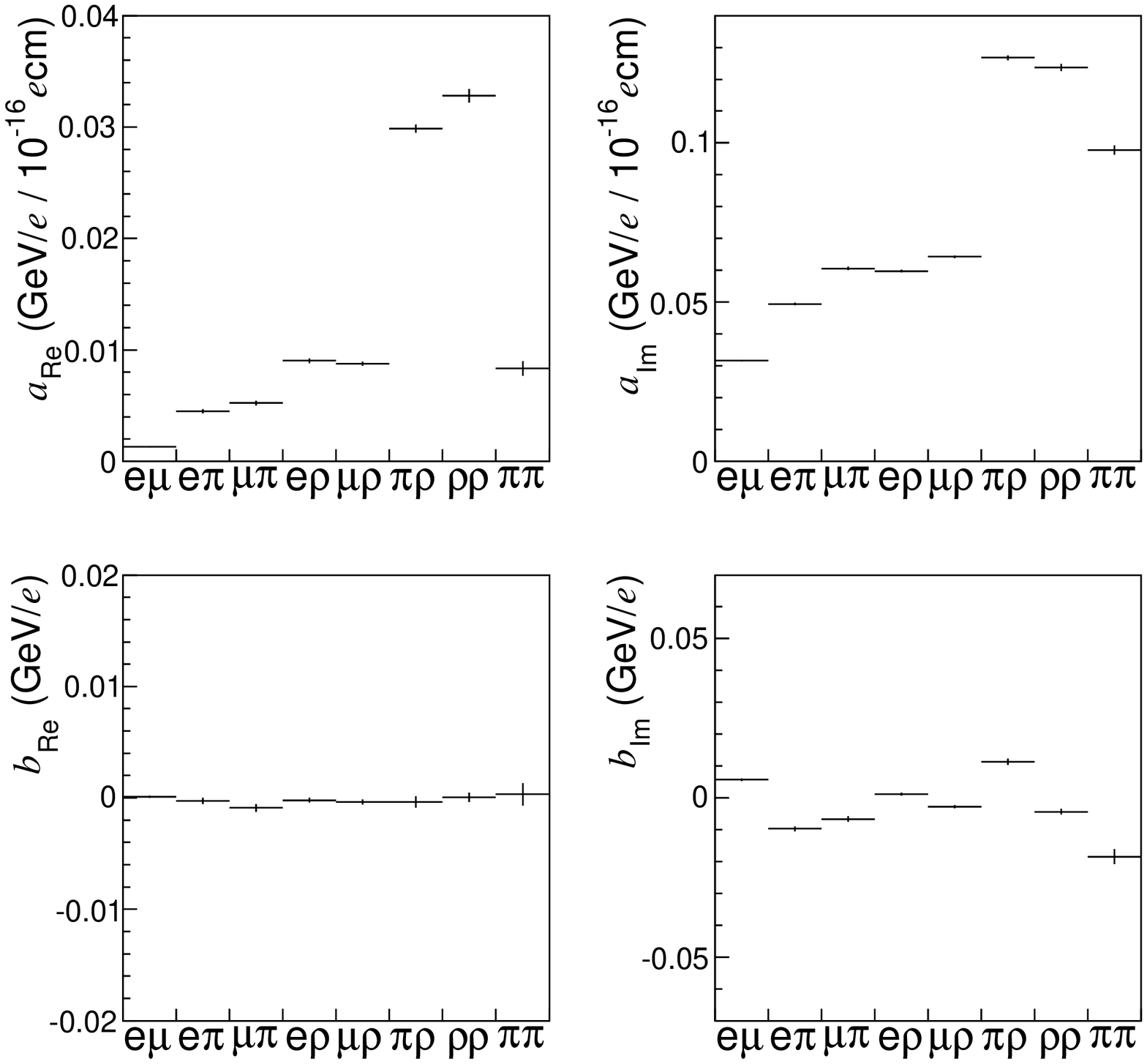}
\caption{EDM parameter sensitivity $a_{\rm Re/Im}$ (top) and offset
$b_{\rm Re/Im}$ (bottom) for each mode.
The uncertainties are due to the statistics of the MC samples.}
\label{fig:coeff}
\end{figure}

%conversion to EDM
To obtain the EDM values from the observables, 
we must determine the relation 
between the EDM and the mean value of the observables
shown in Eq.~(\ref{eq:relation0}).
In order to take into account
the finite detector acceptance, 
the use of the most probable (rather than actual) spin direction,
the ambiguity from the resolution,
the unknown $\tau$ direction, and missing neutrinos,
the relation between the EDM and the mean value of the observables,
$\langle{\cal O}_{\rm Re/Im}\rangle$,
is evaluated using MC simulation for various values of the EDM.
%Figure~\ref{fig:relation} shows the correlation between the EDM value
%and the mean value of the observable for each mode. 
%These relations are obtained from the signal MC samples 
%with the detector simulation and the event selection.
By fitting the relation with a linear function in Eq.~(\ref{eq:relation0}),
as shown in Fig.~\ref{fig:relation},
the coefficients $a_{\rm Re/Im}$ and the offsets $b_{\rm Re/Im}$ are obtained,
which are plotted in Fig.~\ref{fig:coeff}.
As seen from the values of the coefficients, $a_{\rm Re/Im}$,
the $\pi\rho$ and $\rho\rho$ modes have the highest sensitivities for
Re($d_\tau$) and Im($d_\tau$), 
thanks to the high spin analyzing power for $\pi$ and $\rho$ modes.
Nonzero offsets seen for the imaginary part,
$b_{\rm Im}$, are due to a forward-backward asymmetry in the
detector acceptance. The effects of the background are also
taken into account in these coefficients.
The coefficients are corrected for by the purity and the coefficients
obtained using background samples.

We examine a number of possible systematic effects on the EDM 
measurements.
%A number of possible systematic effects have been examined.
The corresponding results are listed 
in Table~\ref{tbl:syserror}.
Differences between the data and simulation 
result in systematic uncertainties. 
%\subsubsection{Detector alignment}
To check for an asymmetry in the tracking systems, 
we analyze $e^+e^- \to \mu^+\mu^-$ events.
We measure the difference of the polar and azimuthal angle of the tracks
between $\mu^+$ and $\mu^-$, 
and then find shifts from the back-to-back direction of 
$-0.67$~mrad for the polar angle and $-0.03$~mrad for 
the azimuthal angle.
By applying an
artificial rotation to one of the charged tracks, 
we obtain residual values of the observables 
and find the results to be less than 10\% of the statistical uncertainties.

%\subsubsection{Momentum reconstruction}

%As shown in the Figs.~\ref{fig:rhomass} and \ref{fig:pi0mass},
There are small data--MC differences in the $\rho$ and $\pi^0$ mass 
distributions.
%between data and MC.
These can be caused by an imperfect momentum reconstruction
resulting in a systematic offset of the observables.
We check the effect of a momentum shift of the charged tracks
by applying a momentum scaling factor of 1.0026,
which is estimated from the peak position of the $\rho$ mass distribution.
We also check the effect of a $\pi^0$ momentum shift
by applying the same factor assuming that the $\rho$ mass difference is 
due to $\pi^0$ momentum shift.
%The result shows that the errors is about 10\% of the statistical ones
%for the modes including $\rho^\pm$.

In the $\pi^0$ invariant mass distribution, % (Figure~\ref{fig:pi0mass}),
we observe a difference of 0.3\% in the mass resolution
between data and MC samples. 
%The difference of the width is 0.3\% of the $\pi^0$ mass.
This is due to data--MC difference in the reconstructed photon energy.
%between data and MC.
We check the effect by changing the photon energies
%according to the $1\sigma$ difference.
according to the difference found in a $D^{*0} \to D^0 \gamma$ study.
The change of the observables is obtained by conservatively
varying the photon energy
to $E_\gamma \pm \sigma$ for both photons from $\pi^0$,
however the change is smaller than the other uncertainties.
%The values $\sigma$ are obtained from 
%the $D^{*0} \to D^0 \gamma$ study.
%Then we found the effect to be about 10\% of the statistical error 
%for the modes including $\rho^\pm$.

%\subsubsection{Efficiency}

The detector response depends on the particle charge, 
especially for electrons and pions.
However, we know that the MC simulation does not exactly reproduce this difference. %The difference in 
%the detection efficiencies brings about an asymmetry of the observables. 
%, because the observables have a charge asymmetry due to the event selection.
We compare the ratios of the yield $N(A^+B^-)/N(A^-B^+)$, where $A$ and $B$
denote the final-state particles, $e$, $\mu$, $\pi$, and $\rho$,
between the data and simulation, and find the difference in ratios to be 
about 1\%.
We apply the observed shift of the charge asymmetry on the yield
to the efficiency, and find
a large systematic uncertainty on the offset of the imaginary part
at the same level as the statistical uncertainty.
(See the entries for Im($d_\tau$) for the $e\pi$ and $\mu\pi$ modes
in Table~\ref{tbl:syserror}.)
The changes in other parameters are negligible.

%\noindent {\bf Efficiency curve} \\
We have checked the polar angle dependence of the charge asymmetry. 
Although the data--MC consistency seems satisfactory,
there are some differences.
We also find a small difference between data and simulation in 
Figs.~\ref{fig:mom} and \ref{fig:cos}, 
where the momentum and polar angle distributions of
the decay product are plotted. %inclusive of their charges.
These differences are probably due to the reconstruction efficiency,
which causes a systematic offset. 
This effect is checked by re-weighting the MC samples with the
weight functions constructed bin-by-bin
from the data--MC ratio for 
the momentum and $\cos \theta$ distributions,
and independently of the charge.
This is the largest source of systematic uncertainty for Re($d_\tau$)
in the high-sensitivity $\pi\rho$ and $\rho\rho$ modes.

%\subsubsection{Background}

In this analysis, the purity is obtained from simulation.
%The difference of the purity between the data and MC leads
%to a systematic error on
%the sensitivities and offsets.
%as shown in eq.(\ref{eq:ab_BGcorrection}).
Any data--MC difference in purity could lead to a bias in the
sensitivities and offsets. 
In order to take into account these possibilities,
we include any difference of yields between the data and simulation
as a systematic uncertainty on the background level.
%We obtain a conservative estimate of the
%systematic uncertainty due to such unknown biases 
%by interpreting the uncertainty in
%the number of background events as a data--MC difference in yield.
The resulting
%Assuming that the error on the number of background events comes from
%the difference of yields between data and MC, the 
systematic uncertainties are about 10\% for the sensitivities 
and about the same order of statistical uncertainties of the observables 
for the offsets.

%\subsubsection{Effect of the initial state radiation}

In addition, 
%we have ignored the effect of the radiations for the 
%calculation of the amplitudes.
we check the effect of ISR by introducing it into the calculation.
We obtain the momenta 
of the ISR photons randomly from the KKMC generator, then,
boost all momenta of the final-state particles into the $\tau$-pair rest frame 
assuming that the ISR is coming from the $e^+ e^-$ beam.
We calculate the observables in this frame.
We iterate this process 100 times using the 
same hit-and-miss approach as in the nominal analysis.
For successful trials, we obtain the mean of the observables.
The shifts and fluctuations with the ISR effect 
give estimates of the systematic effects of ignoring it.

\begin{table*}
\caption{Systematic uncertainties for ${\rm Re}(d_\tau)$ and ${\rm Im}(d_\tau)$
in units of $10^{-17}e\,{\rm cm}$.}
 \label{tbl:syserror}
 \begin{center}
 \begin{tabular}{@{\hspace{0.3cm}}l@{\hspace{0.3cm}}c@{\hspace{0.3cm}}c@{\hspace{0.3cm}}c@{\hspace{0.3cm}}c@{\hspace{0.3cm}}c@{\hspace{0.3cm}}c@{\hspace{0.3cm}}c@{\hspace{0.3cm}}c@{\hspace{0.3cm}}}
  \hline\hline
${\rm Re}(d_\tau)$ &$e\mu$&$e\pi$&$\mu\pi$&$e\rho$&$\mu\rho$&$\pi\rho$&$\rho\rho$&$\pi\pi$\\
  \hline
Detector alignment       & 0.2 & 0.2 & 0.1 & 0.1 & 0.2 & 0.1 & 0.1 & 0.3 \\
Momentum reconstruction  & 0.1 & 0.6 & 0.5 & 0.1 & 0.3 & 0.2 & 0.1 & 1.5 \\
Charge asymmetry         & 0.0 & 0.0 & 0.1 & 0.0 & 0.0 & 0.0 & 0.0 & 0.0 \\
Kinematic dependence of reconstruction efficiency & 3.2 & 4.8 & 3.8 & 0.9 & 2.2 & 0.9 & 0.9 & 3.6 \\
Data--MC diffedence in backgrounds & 1.6 & 0.3 & 1.7 & 0.4 & 0.2 & 0.2 & 0.2 & 3.5 \\
Radiative effects        & 0.7 & 0.5 & 0.6 & 0.2 & 0.2 & 0.0 & 0.0 & 0.1 \\
  \hline
Total                    & 3.6 & 4.8 & 4.3 & 1.0 & 2.2 & 1.0 & 0.9 & 5.2 \\
  \hline \hline
${\rm Im}(d_\tau)$ &$e\mu$&$e\pi$&$\mu\pi$&$e\rho$&$\mu\rho$&$\pi\rho$&$\rho\rho$&$\pi\pi$\\
  \hline
Detector alignment       & 0.0 & 0.0 & 0.0 & 0.0 & 0.1 & 0.0 & 0.0 & 0.0 \\
Momentum reconstruction  & 0.2 & 0.5 & 0.4 & 0.0 & 0.1 & 0.1 & 0.1 & 0.1 \\
Charge asymmetry         & 0.2 & 2.0 & 2.4 & 0.1 & 0.1 & 1.1 & 0.0 & 0.0 \\
Kinematic dependence of reconstruction efficiency & 1.0 & 0.9 & 0.6 & 0.5 & 0.8 & 0.4 & 0.4 & 1.2 \\
Data--MC diffedence in backgrounds     & 1.4 & 0.0 & 0.7 & 0.3 & 0.1 & 0.1 & 0.1 & 0.1 \\
Radiative effects        & 0.1 & 0.1 & 0.1 & 0.1 & 0.1 & 0.0 & 0.0 & 0.0 \\
  \hline
Total                    & 1.8 & 2.2 & 2.6 & 0.6 & 0.8 & 1.2 & 0.4 & 1.2 \\
  \hline\hline
 \end{tabular}
 \end{center}
\end{table*}

%\section{Results}

We calculate the final EDM values using the 833~fb$^{-1}$ data sample,
the results of which are listed in Table~\ref{table:EDMresult}
for each mode.
\begin{table}[htb]
 \begin{center}
 \caption{Results on the $\tau$ electric dipole moment obtained
using 833~fb$^{-1}$ of data.
The first uncertainties are statistical and the second ones are systematic.} 
 \label{table:EDMresult}
 \begin{tabular}{@{\hspace{0.3cm}}c@{\hspace{0.3cm}}r@{\hspace{0.3cm}}r@{\hspace{0.3cm}}}
  \hline\hline
  Mode       & ${\rm Re}(d_\tau) (10^{-17} ~e{\rm cm})$&${\rm Im}(d_\tau) (10^{-17} ~e{\rm cm})$\\
  \hline
      $e\mu$ & $ -3.2 \pm 2.5 \pm 3.6$ & $ 0.6 \pm 0.4 \pm 1.8$ \\
      $e\pi$ & $  0.7 \pm 2.3 \pm 4.8$ & $ 2.4 \pm 0.5 \pm 2.2$ \\
    $\mu\pi$ & $  1.0 \pm 2.2 \pm 4.3$ & $ 2.4 \pm 0.5 \pm 2.6$ \\
     $e\rho$ & $ -1.2 \pm 0.8 \pm 1.0$ & $ -1.1 \pm 0.3 \pm 0.6$ \\
   $\mu\rho$ & $  0.7 \pm 1.0 \pm 2.2$ & $ -0.5 \pm 0.3 \pm 0.8$ \\
   $\pi\rho$ & $ -0.6 \pm 0.7 \pm 1.0$ & $ 0.4 \pm 0.3 \pm 1.2$ \\
  $\rho\rho$ & $ -0.4 \pm 0.5 \pm 0.9$ & $ -0.3 \pm 0.3 \pm 0.4$ \\
    $\pi\pi$ & $ -2.2 \pm 4.3 \pm 5.2$ & $ -0.9 \pm 0.9 \pm 1.2$ \\
  \hline\hline
 \end{tabular}
 \end{center}
\end{table}
%By adding the statistical and systematic errors in quadrature,
We obtain the mean values of the electric dipole moment
weighted by a quadrature sum of statistical and systematic uncertainties,
for the real and imaginary parts,
\begin{eqnarray}
 {\rm Re}(d_\tau) &=& ( -0.62 \pm 0.63 ) \times 10^{-17} ~e{\rm cm}, \\
 {\rm Im}(d_\tau) &=& ( -0.40 \pm 0.32 ) \times 10^{-17} ~e{\rm cm}.
\end{eqnarray}
The 95\% confidence intervals become
\begin{eqnarray}
-1.85 \times 10^{-17} <&{\rm Re}(d_\tau)&< 0.61 \times 10^{-17} ~e{\rm cm}, \\
-1.03 \times 10^{-17} <&{\rm Im}(d_\tau)&< 0.23 \times 10^{-17} ~e{\rm cm}.
\end{eqnarray}
Compared to the previous analysis~\cite{ref:BellePrev}, the
obtained statistical uncertainties are reduced in proportion to the increase in
the data size. The systematic uncertainties are improved because of the improved
simulation, corrections and the larger statistics
of the MC samples.
The sensitivity for Re($d_\tau$) and Im($d_\tau$) has improved by about
a factor of three.
The systematic uncertainty from the detector modeling limits our result
and needs to be controlled for future analysis.

\medskip
\noindent
{\bf Appendix}

The spin vectors used in the analysis are listed here.

For $\tau \to \ell \nu_\ell \nu_\tau$,
\begin{eqnarray}
\Vec{S}_\pm = A
\left( \pm m_\tau \Vec{p}_{\ell^\pm} - 
       \frac{c_\pm + E_{\ell^\pm}m_\tau}{k_0 + m_\tau}\Vec{k}
\right), \\
~~~ A = \frac{4c_\pm -m_\tau^2 -3m_\ell^2}
{3m_\tau^2 c_\pm -4c_\pm^2 -2m_\ell^2 m_\tau^2 +3c_\pm m_\ell^2}, \nonumber \\
c_\pm = k_0 E_{\ell^\pm} \mp \Vec{k}\cdot\Vec{p}_{\ell^\pm}, \nonumber
\end{eqnarray}
where
$k_0$ is the energy of the $\tau^\pm$,
$m_\tau$ is the $\tau$ mass,
$\Vec{k}$ is the three-momentum of the $\tau^+$,
$\Vec{p}_{\ell^\pm}$, $E_{\ell^\pm}$ and $m_\ell$ are the monentum, energy and 
mass of $\ell^\pm$, respectively.

For $\tau \to \rho \nu_\tau \to \pi \pi^0 \nu_\tau$,
\begin{eqnarray}
\Vec{S}_\pm = \mp A
\left( \mp H_0^\pm \Vec{k} + m_\tau \Vec{H}^\pm + 
       \frac{\Vec{k}(\Vec{k}\cdot\Vec{H}^\pm)}{(k_0+m_\tau)}
\right), \\
A = \frac{1}{(k_\pm H^\pm)-m_\tau^2(p_{\pi^\pm}-p_{\pi^0})^2}, \nonumber \\
(H^\pm)^\nu = 2 
 (p_{\pi^\pm}-p_{\pi^0})^\nu (p_{\pi^\pm}-p_{\pi^0})^\mu (k_\pm)_\mu \nonumber \\
~~~~ + (p_{\pi^\pm}+p_{\pi^0})^\nu (p_{\pi^\pm}-p_{\pi^0})^2, \nonumber
\end{eqnarray}
where $k_\pm = (k_0, \pm\Vec{k})$, $H^\pm = (H_0^\pm, \Vec{H}^\pm)$,
and $k_\pm H^\pm$ is the four-vector product. Here,
$p_{\pi^\pm}$ and $p_{\pi^0}$ are the four-momenta of the final state
$\pi^\pm$ and $\pi^0$. 

\medskip
%***** Acknowledgments *****

%\section{Acknowledgements}

We thank the KEKB group for the excellent operation of the
accelerator; the KEK cryogenics group for the efficient
operation of the solenoid; and the KEK computer group, and the Pacific Northwest National
Laboratory (PNNL) Environmental Molecular Sciences Laboratory (EMSL)
computing group for strong computing support; and the National
Institute of Informatics, and Science Information NETwork 5 (SINET5) for
valuable network support.  We acknowledge support from
the Ministry of Education, Culture, Sports, Science, and
Technology (MEXT) of Japan, the Japan Society for the 
Promotion of Science (JSPS) including in particular the
Grant-in-Aid for Scientific Research (A) 19H00682,
and the Tau-Lepton Physics 
Research Center of Nagoya University; 
the Australian Research Council including grants
DP180102629, % Sevior
DP170102389, % Varvell
DP170102204, % Yabsley
DP150103061, % Urquijo
FT130100303; % Urquijo;
Austrian Federal Ministry of Education, Science and Research (FWF) and
FWF Austrian Science Fund No.~P~31361-N36;
the National Natural Science Foundation of China under Contracts
No.~11435013,  %Zhen-An Liu
No.~11475187,  %Chang-Zheng Yuan
No.~11521505,  %Chang-Zheng Yuan
No.~11575017,  %Cheng-Ping Shen
No.~11675166,  %Wen-Biao Yan
No.~11705209;  %Yi-Ming Li
Key Research Program of Frontier Sciences, Chinese Academy of Sciences (CAS), Grant No.~QYZDJ-SSW-SLH011; % Chang-Zheng Yuan
the  CAS Center for Excellence in Particle Physics (CCEPP); %Chang-Zheng Yuan,
the Shanghai Pujiang Program under Grant No.~18PJ1401000;  %Tao Luo
the Shanghai Science and Technology Committee (STCSM) under Grant No.~19ZR1403000; %Xiaolong Wang
the Ministry of Education, Youth and Sports of the Czech
Republic under Contract No.~LTT17020;
Horizon 2020 ERC Advanced Grant No.~884719 and ERC Starting Grant No.~947006 ``InterLeptons'' (European Union);
the Carl Zeiss Foundation, the Deutsche Forschungsgemeinschaft, the
Excellence Cluster Universe, and the VolkswagenStiftung;
the Department of Atomic Energy (Project Identification No. RTI 4002) and the Department of Science and Technology of India; 
the Istituto Nazionale di Fisica Nucleare of Italy; 
National Research Foundation (NRF) of Korea Grant
Nos.~2016R1\-D1A1B\-01010135, 2016R1\-D1A1B\-02012900, 2018R1\-A2B\-3003643,
2018R1\-A6A1A\-06024970, 2018R1\-D1A1B\-07047294, 2019K1\-A3A7A\-09033840,
2019R1\-I1A3A\-01058933;
Radiation Science Research Institute, Foreign Large-size Research Facility Application Supporting project, the Global Science Experimental Data Hub Center of the Korea Institute of Science and Technology Information and KREONET/GLORIAD;
the Polish Ministry of Science and Higher Education and 
the National Science Center;
the Ministry of Science and Higher Education of the Russian Federation, Agreement 14.W03.31.0026, % from 15.02.2018
and the HSE University Basic Research Program, Moscow; % from 15.04.2021
University of Tabuk research grants
S-1440-0321, S-0256-1438, and S-0280-1439 (Saudi Arabia);
the Slovenian Research Agency Grant Nos. J1-9124 and P1-0135;
Ikerbasque, Basque Foundation for Science, Spain;
the Swiss National Science Foundation; 
the Ministry of Education and the Ministry of Science and Technology of Taiwan;
and the United States Department of Energy and the National Science Foundation.

\end{document}